\documentclass[11pt, preprint]{aastex631}
\usepackage{rotating}
\usepackage{bm}
\graphicspath{{./}{figures/}}

\newcommand{\imut}{i_\mathrm{mut}}
\newcommand{\Hill}{\mathrm{Hill}}

\newcommand{\inp}{\mathrm{in}}
\newcommand{\outp}{\mathrm{out}}
\newcommand{\ins}{{12}}
\newcommand{\outs}{{012}}

\usepackage{xcolor}
\definecolor{steelblue}{rgb}{0.275, 0.510, 0.706}
\definecolor{seagreen}{rgb}{0.190, 0.525, 0.361}
\definecolor{alcrimson}{rgb}{0.227, 0.38, 0.54}
\usepackage{xspace}

\received{August 31, 2024}
\revised{April 3, 2025}
\accepted{April 4, 2025}
\submitjournal{ApJ}
\shorttitle{Stability of hierarchical triples in eccentric orbits} \shortauthors{Hayashi, Trani, and Suto}
\begin{document}

\title{Stability of hierarchical triples comprising a central massive body and a tight binary: \\
    the effect of inner and outer eccentricities on the binary breakup condition}
\correspondingauthor{Toshinori Hayashi}
\email{toshinori.hayashi@yukawa.kyoto-u.ac.jp}
\author[0000-0003-0288-6901]{Toshinori Hayashi}
\affiliation{Yukawa Institute for Theoretical Physics, Kyoto
  University, Kyoto 606-8502, Japan}
\author[0000-0001-5371-3432]{Alessandro A. Trani}
\affiliation{Niels Bohr Institute, University of Copenhagen,
  Blegdamsvej 17, 2100 Copenhagen, Denmark}
\author[0000-0002-4858-7598]{Yasushi Suto}
\affiliation{Research Institute, Kochi University of Technology, Tosa
  Yamada, Kochi 782-8502, Japan}
\affiliation{Department of Physics,
  The University of Tokyo, Tokyo 113-0033, Japan}
\affiliation{Research Center for the Early Universe, School of
  Science, The University of Tokyo, Tokyo 113-0033, Japan}

\begin{abstract}
	We explore the stability of gravitational triple systems
  comprising a central massive body and a tight binary of less massive
  pairs. In the present paper, we focus on improving the Hill-type stability criterion for the binary in those systems, with particular attention to the effects of the eccentricities of the inner and outer orbits. We perform direct Newtonian N-body simulations over much longer integration times than
  previous studies, which is essential to determine the stability and breakup timescale distributions of those systems in a reliable fashion. As a result, we obtain an empirical fitting formula of the binary breakup condition that
  incorporates effects of the inner and outer eccentricities, the
  mutual inclination of the inner and outer orbits, the mass ratios of
  the three bodies, and the breakup timescale.
\end{abstract}

\section{Introduction \label{sec:intro}} 

The stability and fate of gravitational triple systems is one
  of the long-standing and challenging questions in mathematical
  physics and astronomy. Many previous authors have approached the
  problem using a variety of different approximations and
  methodologies, including
  \cite{Eggleton1995,Holman1999,Mardling1999,Mardling2001,
    Georgakarakos2013,Grishin2017,He2018,
    Myllari2018,Wei2021,Lalande2022,Tory2022,Vynatheya2022,HTS2022,HTS2023,Zhang2023},
  among others.

In particular, a lot of attention has been paid to a
  hierarchical triple configuration, in which two of them form a tight
  binary and interact with the tertiary object.  For instance,
  \citet{Mardling2001} (hereafter, MA01) considered their stability
  combining the chaos theory and direct numerical simulations, and
  obtained the following stability criterion:
\begin{equation}
  \label{eq:MA-criterion}
  \frac{a_\outp(1-e_\outp)}{a_\inp}
	> 2.8\left(1-0.3\frac{\imut}{180^\circ}\right)
	\left[\left(1+ \frac{m_3}{m_1+m_2}\right)
	\frac{(1+e_\outp)}{\sqrt{1-e_\outp}}\right]^{2/5}.
\end{equation}
In the above expression, $a_\outp$ and $a_\inp$ are the semi-major axes of
the outer and inner orbits, $e_\outp$ is the orbital eccentricity of the
outer orbit, $\imut$ is the mutual inclination between the outer and
inner orbits, $m_3$ is the mass of the tertiary, and $m_1$ and $m_2$
are the masses of the inner binary.

The criterion (\ref{eq:MA-criterion}) (implicitly) assumes that the
tertiary mass $m_3$ is at most comparable to $m_1$ and $m_2$, and has
been frequently applied to the case of $m_3 \ll m_{12} \equiv
m_1+m_2$.  In what follows, we refer to such configurations as HT-P
(Hierarchical Triple Planet-type) in which a tertiary orbits an inner
massive binary. The classic criterion (\ref{eq:MA-criterion}) has been
tested and improved recently by \cite{Vynatheya2022,HTS2022,HTS2023},
for example.

Another hierarchical triple configuration, which we refer to as
  HT-S (Hierarchical Triple Satellite-type) below, is a central
  massive object orbited by a less massive binary. The dynamics of
  HT-S configurations is important in understanding the fate of a
  variety of physically interesting triple systems.  For instance, the
  enhanced merging rate of Binary Black-Holes (BBHs) around
  supermassive BH (SMBH) \citep[e.g.][]{Li2015, VanLandingham2016} may
  be a major target of future space-based gravitational wave detectors
  \citep[e.g.][]{Xuan2023}.  Extrasolar binary-planet systems also
  belong to this configuration, which have been theoretically
  predicted/discussed \citep[e.g.][]{Ochiai2014,Lewis2015}, but not
  yet detected.

The present paper considers the stability of HT-S systems comprising a central massive body of mass $m_0$ orbited by an initially tightly-bound binary of masses $m_1$ and $m_2$. We examine the breakup condition of a binary, and obtain the empirical formulae of breakup timescales as functions of the initial parameters of systems. To avoid confusion, we use $m_0$ for the massive tertiary in HT-S systems, while we use $m_3$ for the less massive tertiary in the HT-P systems.

Our current analysis assumes purely Newtonian gravity for simplicity and clarity, and the effect of general relativity will be studied in a forthcoming paper. The subsequent evolution and fate after binary breakups are also important to fully understand HT-S stability, keeping in mind the application to black-hole triples. Nevertheless, we focus on the breakup condition in the present paper, and the subsequent evolution will be separately discussed elsewhere.

The Hill stability condition is particularly useful in
  considering the stability of HT-S systems.  Historically, the Hill
stability was first derived for the motion of moon, applying the
conservation of the Jacobi integral in restricted three-body problems
\citep[e.g.][]{Hill1878}. Later, it was extended to more general
(non-restricted) three-body problems using Sundman's inequality, which determines a required condition among momenta for a gravitational multi-body system,  
and the sufficient conditions for some cases were derived
\citep[e.g.][]{Marchal1982}.

While the Hill stability is
  rigorously defined in those papers, the condition is roughly
  understood as the competition between the gravitational tidal force
  of the binary due to the central object and the mutual gravitational
  attraction of the binary pairs.

Consider two objects of masses $m_1$ and $m_2$ orbiting a more massive
  central object of mass $m_0$. If both the inner and outer orbits are circular, under the test particle limit ($m_2
  \ll m_1 < m_0$), a binary companion $m_2$ orbiting $m_1$
  is stably bound to $m_1$ if its semi-major axis $a_\ins$ is less
  than the Hill radius $a_\Hill$:
\begin{equation}
  \label{eq:Hill-stability-a}
  a_\ins < a_\Hill \equiv a_\outs \left(\frac{m_1}{3m_0}\right)^{1/3}.
\end{equation}
where $a_\outs$ is the semi-major axis between $m_1$--$m_2$ binary and $m_0$. The above stability condition is
rewritten in terms of the inner and outer orbital periods as
\begin{equation}
  \label{eq:Hill-stability-Period}
  {P_\outs} >  \sqrt{3} {P_\ins}.
\end{equation}
where $P_\outs$ and $P_\ins$ are the periods of outer and inner orbits,
respectively. In other words, the Hill-stability is roughly equivalent to
the condition that the outer orbital period is longer than the inner
one. Note that, for clarity, we use the labels `$\ins$' and `$\outs$' for
HT-S, instead of `$\inp$' and `$\outp$' of HT-P, unless otherwise
specified.

While the Hill stability condition (\ref{eq:Hill-stability-a})
  or (\ref{eq:Hill-stability-Period}) gives approximately the
  stability condition for the HT-S systems, it needs to be generalized
  to the cases of non-circular ($e_{12} \not=0$ and/or $e_{012}
  \not=0$) orbits, the finite mass of $m_2$, and non-coplanar orbits.

For instance, \citet{Grishin2017} (hereafter GPZM17)
  generalized the stability condition (\ref{eq:Hill-stability-a}) by
  considering the mutual inclination dependence for initially near
  circular inner and outer orbits:
\begin{eqnarray}
  \label{eq:Grishin-criterion}
  & \displaystyle \frac{a_\outs}{a_\ins}
   >  \frac{1}{3^{1/3}} 
   \left(\frac{3m_0}{m_1+m_2}\right)^{1/3}
  \left(\cos\imut + \sqrt{3+\cos^2\imut}\right)^{2/3} \cr
& \hspace{7cm} \times  \left\{
\begin{array}{ll}
1  &\qquad (\cos^2{\imut} > 3/5)\\
\displaystyle
\frac{9-5\cos^2{\imut}}{6} & \qquad (\cos^2{\imut} \leq 3/5)
\end{array}
\right.
.
\end{eqnarray}

The factor $ \left(\cos\imut +
  \sqrt{3+\cos^2\imut}\right)^{2/3}$ in equation (\ref{eq:Grishin-criterion})
  was derived by GPZM17 as follows. As discussed in
  \citet{Innanen1980}, the acceleration $\bm{a}$
  acting on a massless tertiary is given by
\begin{eqnarray}
  \bm{a} = \ddot{\bm{R}}_{02} + \dot{\bm{\Omega}}_{01} \times \bm{r}_\ins
  + \bm{\Omega}_{01} \times (\bm{\Omega}_{01} \times \bm{r}_\ins)
  + 2\bm{\Omega}_{01} \times \bm{v}_r ,
\end{eqnarray}
where $\bm{R}_{02}$, $\bm{r}_\ins$, $\bm{\Omega}_{01}$ and
$\bm{v}_r$ are the distances of the tertiary from $m_0$ and $m_1$, the
angular velocity of $m_1$ around $m_0$, and the velocity of the tertiary,
respectively. Assuming the force balance $\bm{a}=\bm{0}$, one
  can obtain the $\imut$ dependence at the boundary.  Furthermore,
  GPZM17 considered the secular effect by the von
  Zeipel-Kozai-Lidov(ZKL) oscillations
  \citep{Zeipel1910,Kozai1962,Lidov1962}, and found an additional
  dependence of $\imut$, the final factor in equation
  (\ref{eq:Grishin-criterion}), by replacing $a_\ins$ with $a_\ins
  (1+0.5e^2_\mathrm{max})$, where $e_\mathrm{max}$ is the maximum
  value of $e_\ins$ attained during the ZKL cycle.  Numerical
  simulations by GPZM17 confirmed that the condition
  (\ref{eq:Grishin-criterion}) well captures the strong
  destabilization of HT-S systems with near-polar orbits.

The present paper aims to further generalize the condition
(\ref{eq:Grishin-criterion}) for the HT-S systems of initially
non-circular orbits ($e_\ins \not=0$ and $e_\outs \not=0$), as
pioneered by MA01 for the HT-P systems. For the dependence of
  $\imut$, we adopt the analytically derived factor $ \left(\cos\imut +
  \sqrt{3+\cos^2\imut}\right)^{2/3}$ in GPZM17, but add an empirical
  correction $h(e_\outs, \imut; T_\mathrm{break}/P_\ins)$ defined
  below. In doing so, we also empirically derive the dependence on
  $e_\ins$ and $e_\outs$ using numerical simulations. The resulting
  approximate empirical corrections, equation (\ref{eq:Upsilon}),
  improve the breakup condition (\ref{eq:Grishin-criterion}) for the HT-S systems.

\begin{figure*}
\begin{center}
\includegraphics[clip,width=11cm]{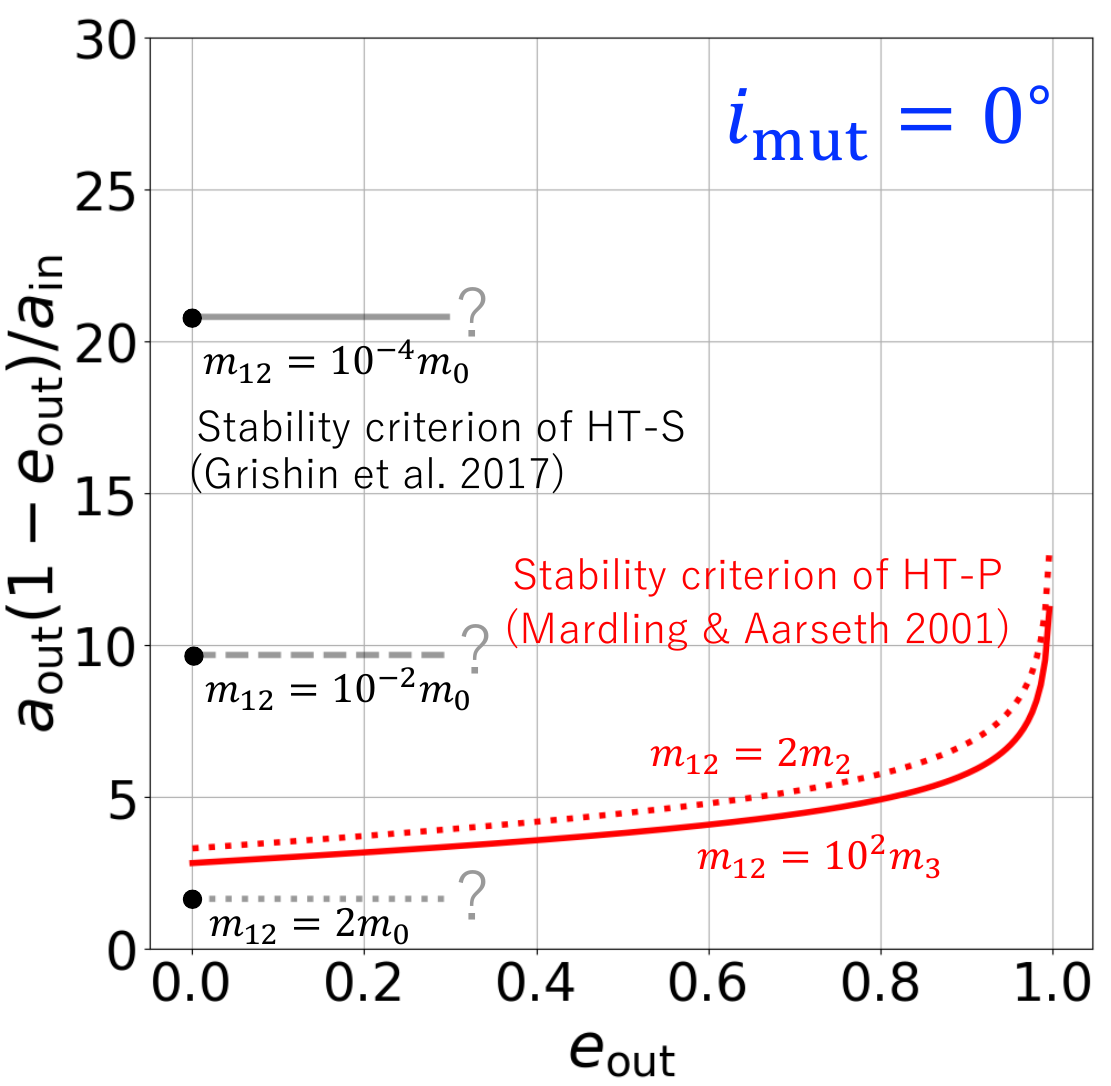}
\end{center}
\caption{Comparison of stability criteria for HT-P ($m_{12}/m_3 = 2,
  10^{2}$, red curves) and HT-S ($m_{12} /m_0= 2, 10^{-2},10^{-4}$,
  black dots) in $e_\outp$--$ a_\outp(1-e_\outp)/a_\inp$
  plane, where $m_{12}=m_1+m_2$. For HT-S, $a_\inp$, $a_\outp$ and $e_\outp$ are interpreted as 
  $a_\ins$, $a_\outs$ and $e_\outs$, respectively. Note that
  the stability criterion of HT-P (GPZM17) focuses on circular
  systems, and horizontal gray lines are simple extrapolations assuming $e_\outs$
  dependence is all absorbed in $a_\outs(1-e_\outs)$.
  \label{fig:P-S-comparison}}
\end{figure*}

In order to clarify our purpose of the present paper, we show the comparison of stability criteria for HT-S and HT-P in
  Figure \ref{fig:P-S-comparison}: the MA01 stability condition
  (\ref{eq:MA-criterion}) for the coplanar HT-P systems in red curves,
  and the GPZM17 stability condition for HT-S systems in black dots.
Because GPZM17 focused on initially circular HT-S systems, their
result (\ref{eq:Grishin-criterion}) is valid for $e_\outs=0$
alone. Incidentally, Figure \ref{fig:P-S-comparison} assumes that the
initial inner orbits are circular ($e_\inp(=e_\ins)=0$), and shows the
mass dependence using different line types for comparison.

If the Hill stability is purely determined by the amplitude of the
instantaneous tidal force due to the central massive body, one naively
expects that the non-circular effect would be incorporated simply by
replacing the initial semi-major axes $a_\ins$ and $a_\outs$ with
their initial apocenter distance $a_\ins (1+e_\ins)$, and pericenter
distance $a_\outs(1-e_\outs)$, respectively. In reality, however, this
simple procedure completely neglects the evolution of the orbital
elements. This is why we explore the non-circular effect using a
series of systematic numerical simulations below. 

\begin{figure*}
\begin{center}
\includegraphics[clip,width=9cm]{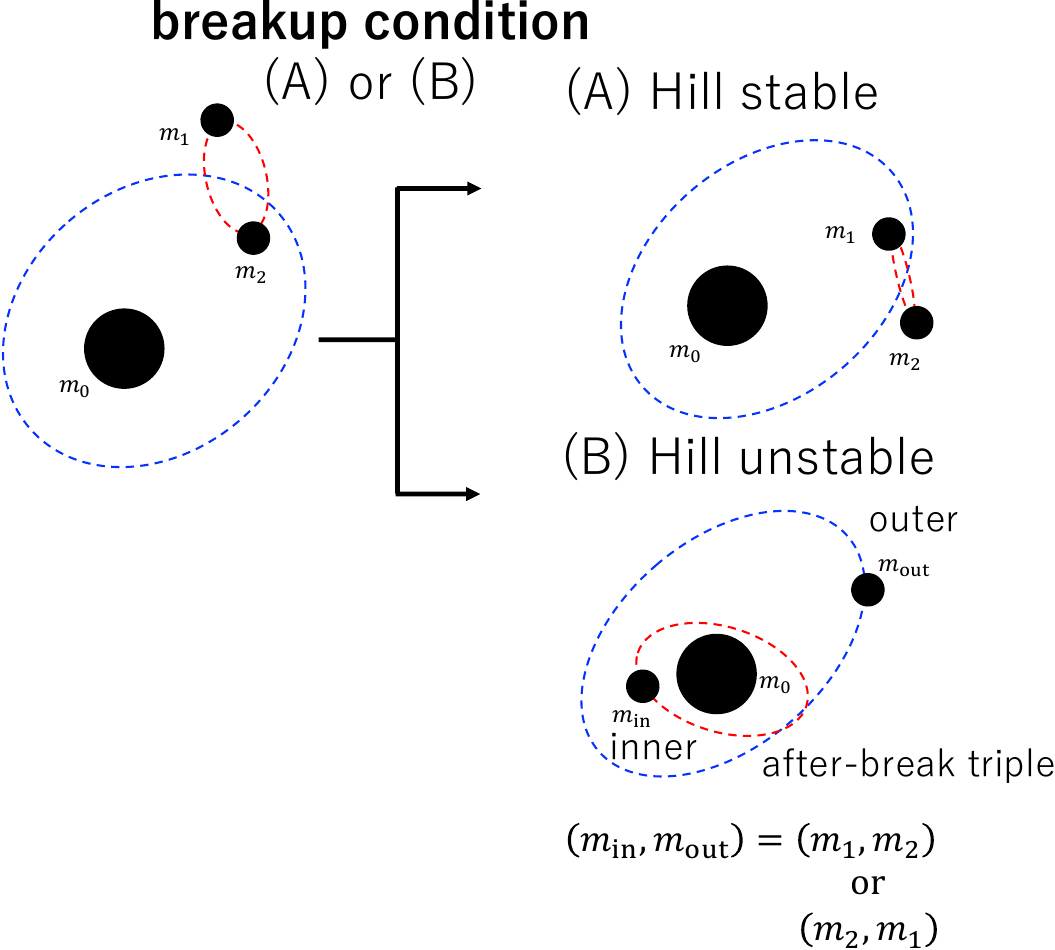}
\end{center}
\caption{Schematic illustration of the evolution of HT-S systems. We check the breakup condition of them: (A) Hill stable or (B) unstable. While the evolution of Hill stable triples (A) can be computed using secular perturbation, that of Hill unstable ones (B) generally require direct N-body simulations.\label{fig:schematic}}
\end{figure*}

Figure \ref{fig:schematic} schematically summarizes the evolution of HT-S systems and the purpose of present paper. We attempt to extend the GPZM17 stability criterion by including initial eccentricity, mutual inclination, and mass ratio, and breakup timescale dependences using long-term N-body integrations. Here, we assume point-mass objects and study their dynamical stability under Newtonian gravity.

We emphasize that the direct N-body simulations, instead of the orbit-averaged secular simulations, are essential to explore the Hill-stability boundary between (A) and (B) of Figure \ref{fig:schematic}. We practically perform the simulations up to $10^8 P_\ins$ throughout this paper, much longer timescale than previous studies, and determine the breakup timescale dependences on the breakup condition.

The rest of the paper is organized as follows. In section
\ref{sec:method}, we first describe initial setup of numerical
simulations, and how we evaluate the breakup condition of initial binary. Then, section \ref{sec:result} shows the resulting breakup criterion, including initial eccentricity, mutual
inclination, mass ratio and breakup time dependences. We also systematically check
the initial phase dependence about the breakup condition, and show the result in the appendix. Finally, section \ref{sec:summary}
summarizes the conclusion of this paper, and discusses possible future prospects.

\section{Method \label{sec:method}}

We carry out a series of N-body simulations using {\tt TSUNAMI}
\citep[see][]{Trani2023}, and explore the parameter space for the
initial values of $e_\ins$ and $e_\outs$ in particular. We also
  consider a set of different mass ratios for the three bodies, and
  different initial mutual inclinations. Those values will be
  specified when we present the results.

 We determine the breakup conditions of those systems on $e_\outs$ -- $a_\outs(1-e_\outs)/a_\ins$ plane, following \citet{HTS2022,HTS2023}. For a given set of values for $e_\ins$, $e_\outs$, $a_\ins$, $\imut$, $m_1$, $m_2$, and $m_0$, we start a simulation with a sufficiently small value of  $a_\outs(1-e_\outs)/a_\ins$ ($= 30$) so that the $m_1$--$m_2$ binary quickly breaks due to the Hill instability. The results indeed show quick breakups, and justify this specific choice. On the other hand, it is impractical to set the upper limit of $a_\outs(1-e_\outs)/a_\ins$ beforehand, since it should highly depend on initial conditions. Therefore, we rather adopt the following procedure.

 We record the breakup time $T_\mathrm{break}$ of an $m_1$--$m_2$ binary for each simulation. However, we practically set the integration time limit $t_\mathrm{int}(=10^8P_\ins)$ and rather classify a system into `unresolved' if the binary does not break within the limit. We stop a simulation either when a binary breaks up, or when a system is classified into `unresolved'. After the simulation, we gradually increase the value of $a_\outs(1-e_\outs)/a_\ins$ by $2$, and run the next simulation.

The above procedure is repeated unless a system meets `unresolved'. If a system satisfies `unresolved', we check a larger value of $a_\outs(1-e_\outs)/a_\ins$ to avoid a possible spurious result due to the fluctuation. If the binary breaks, we again continue increasing $a_\outs(1-e_\outs)/a_\ins$, but if two consecutive simulations satisfy `unresolved', we stop the simulation sequence for a given set of initial parameters.

Although there may be possible fine structures above the determined boundary (islands of instability due to resonances, for instance), we are interested in the global stability boundary, and we check the robustness of our derived boundary by running realizations with different initial phases later (see the appendix).

 All the above parameter survey is done for a given set of the
  initial phases (mean anomalies and pericenter arguments), but the
  initial-phase dependence is checked later for a sample of specific
  orbital configurations. Throughout this paper, we fix $P_\ins$ as $1$
  yr unless otherwise stated, but this does not virtually affect our
  result thanks to the scaling relation in Newtonian gravity.

In this procedure, the definition of the binary breakup is the
  most crucial, and we basically apply the disruption condition in
  \citet{HTS2023} as follows. At each numerical timestep during the
  simulation, we compute the osculating Kepler orbital elements for
  bounded pairs of particles; $(m_1,m_2)$, $(m_0,m_1)$ and
  ($m_0,m_2$) at most. Then, we define that the initial binary pair
  $(m_1,m_2)$ is broken if 
  \begin{equation}
    \label{eq:binary-break}
  0 < a_{01}(t) < a_{12}(t) \qquad \mathrm{or} \qquad 0 < a_{02}(t) < a_{12}(t)  \qquad \mathrm{or} \qquad a_{12}(t) < 0.
\end{equation}
The above conditions basically reflect the configuration of (B) in Figure \ref{fig:schematic}. We adopt the semi-major axes, instead of the binary
binding energies, in the above definition of the binary break. This is
because the amplitude of binding energies is always dominated by the
central massive body in the present case ($m_0 \gg m_1 \approx m_2$).
    
GPZM17 adopted a binary breakup criterion when either (1) the eccentricity
  of a binary exceeds unity or (2) the distance of binary exceeds
  $3a_\mathrm{Hill}$. We prefer the condition (\ref{eq:binary-break}),
  rather than that adopted by GPZM17 for the following reasons.
  First, the osculating eccentricity may not be stable especially when the
  binary becomes close to a breakup configuration.  Second, to use the
  Hill-stability radius as a measure of the breakup may lead to a
  circular argument, because our current purpose is to find the
  improved criterion for the Hill stability. In practice, however, we numerically made sure that the two
  different definitions change the stability boundary, {\it i.e.,}
  the value of the y-axis in Figure \ref{fig:P-S-comparison}, usually
  within tens of percent as shown in the next section.

\section{Improved formula for breakup conditions of the HT-S systems \label{sec:result}}
In this section, we derive an empirical formula for breakup condition of the HT-S system that improves the GPZM17 stability
criterion (\ref{eq:Grishin-criterion}). We examine the
  dependence on the inner and outer eccentricities for eight different
  cases of the mutual inclinations: $\imut = 30n^\circ$($n=0$--$6$),
  and $\imut = 5^\circ$ (near-coplanar-prograde case). We just include $\imut=5^\circ$ in addition to $0^\circ$ to avoid possible specificity of a complete coplanar system although the result turns out to be almost identical.

  Furthermore, for prograde examples, we confirmed the
$m_0/m_{12}$--dependence in the GPZM17 criterion is valid for a wide
range of $m_0/m_{12}$, and make sure that the result is statistically
robust against the initial orbital phases of the three bodies (see the
appendix). Because the breakup condition is by nature dependent
  on the integration time, we consider also the time dependence of the breakup conditions. We note that binary breakup time $T_{\rm break}$
  is evaluated in units of the initial value of $P_\ins$ ($1$ yr in the present paper), without loss of generality due to the scale invariance under  Newtonian gravity.

As specified in
  introduction, we rely on the analytically derived $\imut$-dependence
  discovered by GPZM17, but we include an empirical correction as
  $h(e_\outs, \imut; T_\mathrm{break}/P_\ins)$. Our final
  stability criterion is written as follows:
{\begin{equation}
	\label{eq:Upsilon}
	\Upsilon \equiv \frac{\tilde{r}_\outs}{\tilde{r}_\ins} >
        \Upsilon_\mathrm{crit} \equiv
	\left(\frac{m_0}{m_1+m_2}\right)^{1/3}
	\left(\cos\imut+\sqrt{3+\cos^2\imut}\right)^{2/3}h(e_\outs, \imut; T_\mathrm{break}/P_\ins),
\end{equation}
where 
\begin{eqnarray}
  \tilde{r}_\outs &\equiv& a_\outs (1-e_\outs), \\
\label{eq:rin-tilde}
\tilde{r}_\ins &=& \left\{
\begin{array}{ll}
   a_\ins (1+0.5e_\ins^2)
  & \quad (\mathrm{no~ZKL~\&~prograde})\\
   a_\ins [1+0.5e_\mathrm{max}^2(\imut)]
  & \quad (\mathrm{ZKL}) \\
  a_\ins (1+e_\ins)
  & \quad (\mathrm{no~ZKL~\&~retrograde}) 
\end{array}
	\right. , 
\end{eqnarray}
and $e_\mathrm{max}$ is the maximum binary eccentricity $e_\ins$
under the ZKL oscillations.

We use the classification based on physical conditions in equation (\ref{eq:rin-tilde}). In the present paper, $(0^\circ$, $5^\circ$, $30^\circ)$, $(60^\circ$, $90^\circ$, $120^\circ)$, and $(150^\circ, 180^\circ)$ correspond to (no ZKL \& prograde), (ZKL), and (no ZKL \& retrograde), respectively. Note that we consider a system in the
  ZKL regime using test-particle approximation, {\it i.e.}
  $\cos^2{\imut} \leq 3/5$, for simplicity.
\begin{equation}
  e_\mathrm{max} = \sqrt {1-\frac{5}{3}\cos^2{\imut}}
  \qquad (\cos^2{\imut} \leq 3/5), 
  \label{eq:emax}
\end{equation}
which recovers the part of the $\imut$-dependence in the GPZM17 criterion
(\ref{eq:Grishin-criterion}):
\begin{equation}
  1+0.5e_\mathrm{max}^2(\imut)
 = \frac{9-5\cos^2{\imut}}{6} .
\end{equation}

The empirical correction factor $h(e_\outs, \imut;T_\mathrm{break}/P_\ins)$ is determined by fitting as follows. For each set of initial parameters, we first divide all the data $(e_\outs, e_\ins, \imut, \tilde{r}_\outs/\tilde{r}_\ins, T_\mathrm{break}/P_\ins)$ (see Figure \ref{fig:imut_norm} also) into $10^2 \leq T_\mathrm{break}/P_\ins < 10^4$, $10^4 \leq T_\mathrm{break}/P_\ins < 10^6$, and $10^6 \leq T_\mathrm{break}/P_\ins < 10^8$ bins. If a bin is empty, we use the minimum $(\tilde{r}_\outs/\tilde{r}_\ins)_\mathrm{norm}$ of the upper bin, where
\begin{equation}
(\tilde{r}_\outs/\tilde{r}_\ins)_\mathrm{norm} \equiv \frac{(\tilde{r}_\outs/\tilde{r}_\ins)}{	\displaystyle{\left(\frac{m_0}{m_1+m_2}\right)^{1/3}}
	\displaystyle{\left(\cos\imut+\sqrt{3+\cos^2\imut}\right)^{2/3}}}. \label{eq:norm}
\end{equation}
Then, we fit the binned data using a simple liner function:
\begin{equation}
h(e_\outs, \imut; T_\mathrm{break}/P_\ins) = C_1(\imut, T_\mathrm{break}/P_\ins) e_\outs + C_2(\imut, T_\mathrm{break}/P_\ins),
\label{eq:h_func}
\end{equation}
where $C_1(i_\mathrm{mut}, T_\mathrm{break}/P_\ins)$ and $C_2(i_\mathrm{mut}, T_\mathrm{break}/P_\ins)$ are fitting parameters. For each choice of $\imut$ and $T_\mathrm{break}/P_\ins$, we obtain the best-fit values of $C_1(i_\mathrm{mut}, T_\mathrm{break}/P_\ins)$ and $C_2(i_\mathrm{mut}, T_\mathrm{break}/P_\ins)$.

The resulting best-fit values are plotted in Figure \ref{fig:fit}, and the specific values are included in Figure \ref{fig:imut_norm}. Figure \ref{fig:fit} shows somewhat further $\imut$ corrections for $\left(\cos\imut+\sqrt{3+\cos^2\imut}\right)^{2/3}$, especially for highly inclined systems.  Here, we note that $Y$ is usually defined as
$a_\outp(1-e_\outp)/a_\inp(1+e_\inp)$ \citep[e.g.][]{Eggleton1995,Vynatheya2022}, and we define $\Upsilon$ as the extension of this quantity.

\begin{figure*}
	\begin{center}
		\includegraphics[clip,width=12cm]{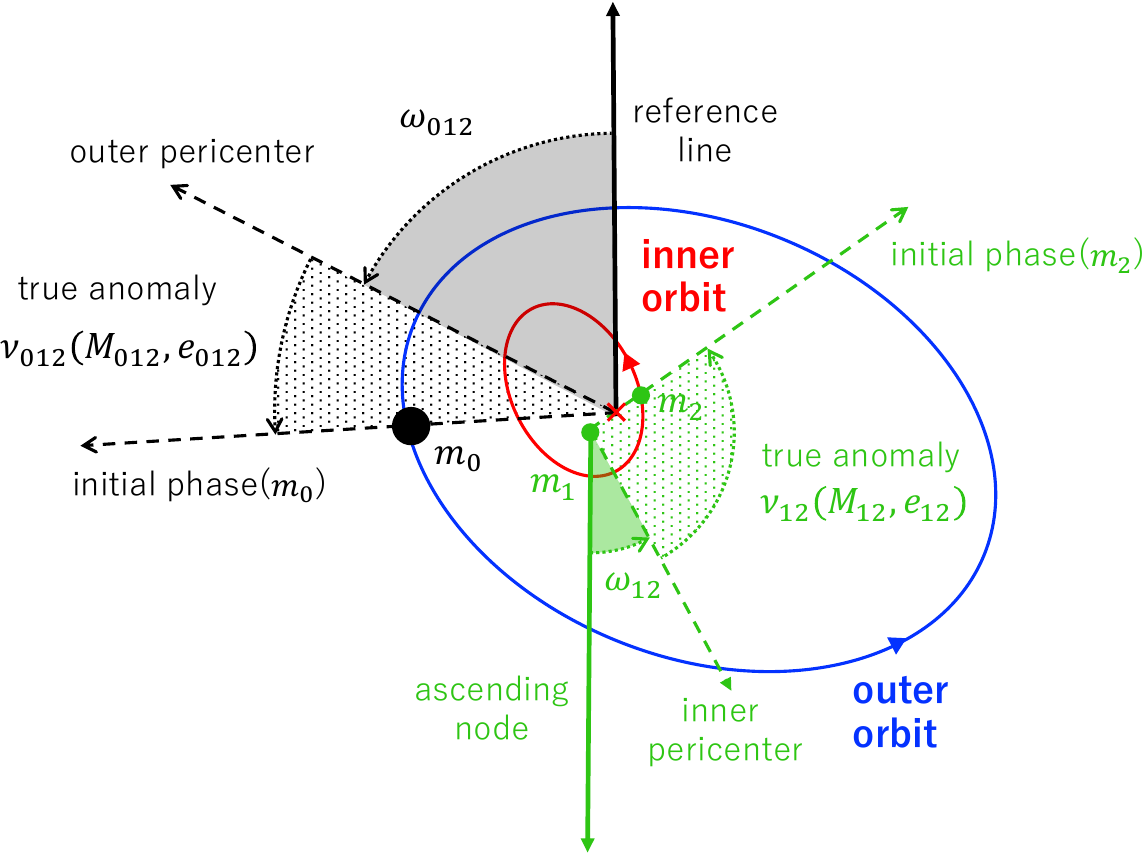}
	\end{center}
	\caption{The initial configuration of the inner and outer
		orbits. 
		\label{fig:orbit}}
\end{figure*}

\begin{figure*}
	\begin{center}
		\includegraphics[clip,width=14cm]{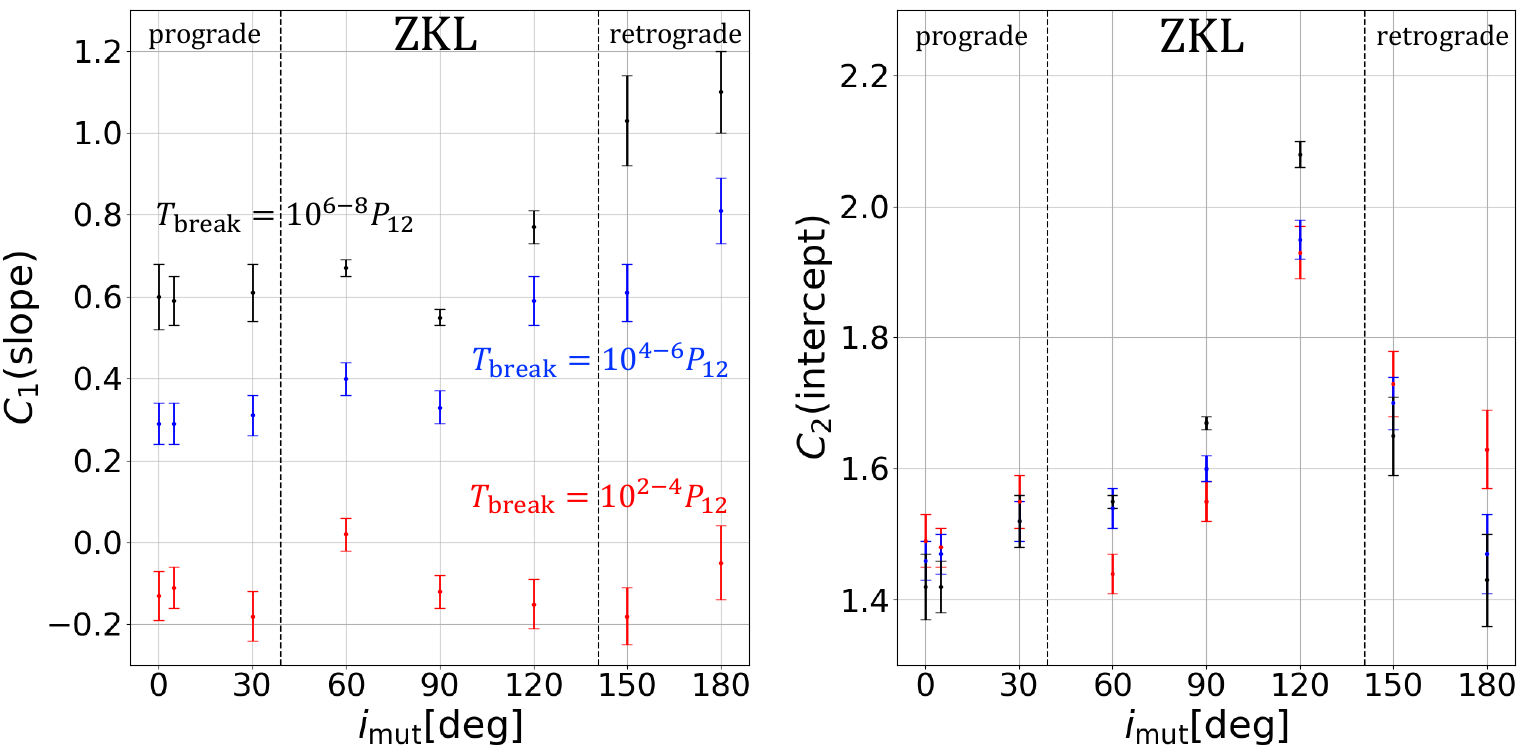}
	\end{center}
	\caption{The best-fit values of $C_1(\imut, T_\mathrm{break}/P_\ins)$ (slope) and $C_2(\imut, T_\mathrm{break}/P_\ins)$ (intercept) in $h(e_\outs, \imut;T_\mathrm{break}/P_\ins)$ (see equation (\ref{eq:h_func})) with $1$--$\sigma$ error bars. Red, blue, and black circles correspond to the results for $T_\mathrm{break}/P_\ins=10^{2-4}$, $10^{4-6}$, and $10^{6-8}$, respectively. For reference, we plot the vertical dashed lines to specify the ZKL regime boundaries according to the test-particle approximation. \label{fig:fit}}
\end{figure*}

The next subsections will describe how we obtain the parameter
dependence in the breakup criterion (\ref{eq:Upsilon}).  We
run a series of simulations for different $\imut$ ($0^\circ$,
  $5^\circ$, $30^\circ$, $60^\circ$, $90^\circ$, $120^\circ$,
  $150^\circ$, and $180^\circ$) fixing initial phases
  $(\omega_\ins,\omega_\outs,$$M_\ins,M_\outs)$ as $(180^\circ,
  0^\circ, 30^\circ, 45^\circ)$, where $M_\ins$, $M_\outs$,
  $\omega_\ins$, and $\omega_\outs$ are the inner mean anomaly, outer
  mean anomaly, inner pericenter argument, and outer pericenter
  argument, respectively (see Figure \ref{fig:orbit}).  We use the
  Jacobi coordinate system, and the invariant plane, therefore fixing
  $\Omega_\ins$ and $\Omega_\outs$ as $180^\circ$ and $0^\circ$,
  respectively.

We examine the dependence of the stability boundary on the
eccentricities $e_\ins$ and $e_\outs$ in \S \ref{subsec:ein} and \S \ref{subsec:eout} for $m_0=10^6M_\odot$ and $m_1=m_2=10~M_\odot$, with
a triple system comprising a massive BH and a stellar mass binary
black hole in mind. Since we neglect the effect of general relativity here, the result is dependent on their mass ratio alone. For prograde examples, we examine the
dependence on their mass ratio in \S
\ref{subsec:massratio}, and the sensitivity to the initial
phases is discussed in appendix \ref{appendix:initialphases}.
\subsection{Inner eccentricity dependence
\label{subsec:ein}}

For the initial configuration described in the above, we examine 25
models with $e_\outs=0, 0.2, 0.4, 0.6, 0.8$, and $e_\ins=0, 0.2, 0.4,
0.6, 0.8$. For each model, we determine the breakup condition following \S \ref{sec:method}, adopting $t_\mathrm{int}=10^8P_\ins$. 
\begin{figure*}
\begin{center}
	\includegraphics[clip,width=14cm]{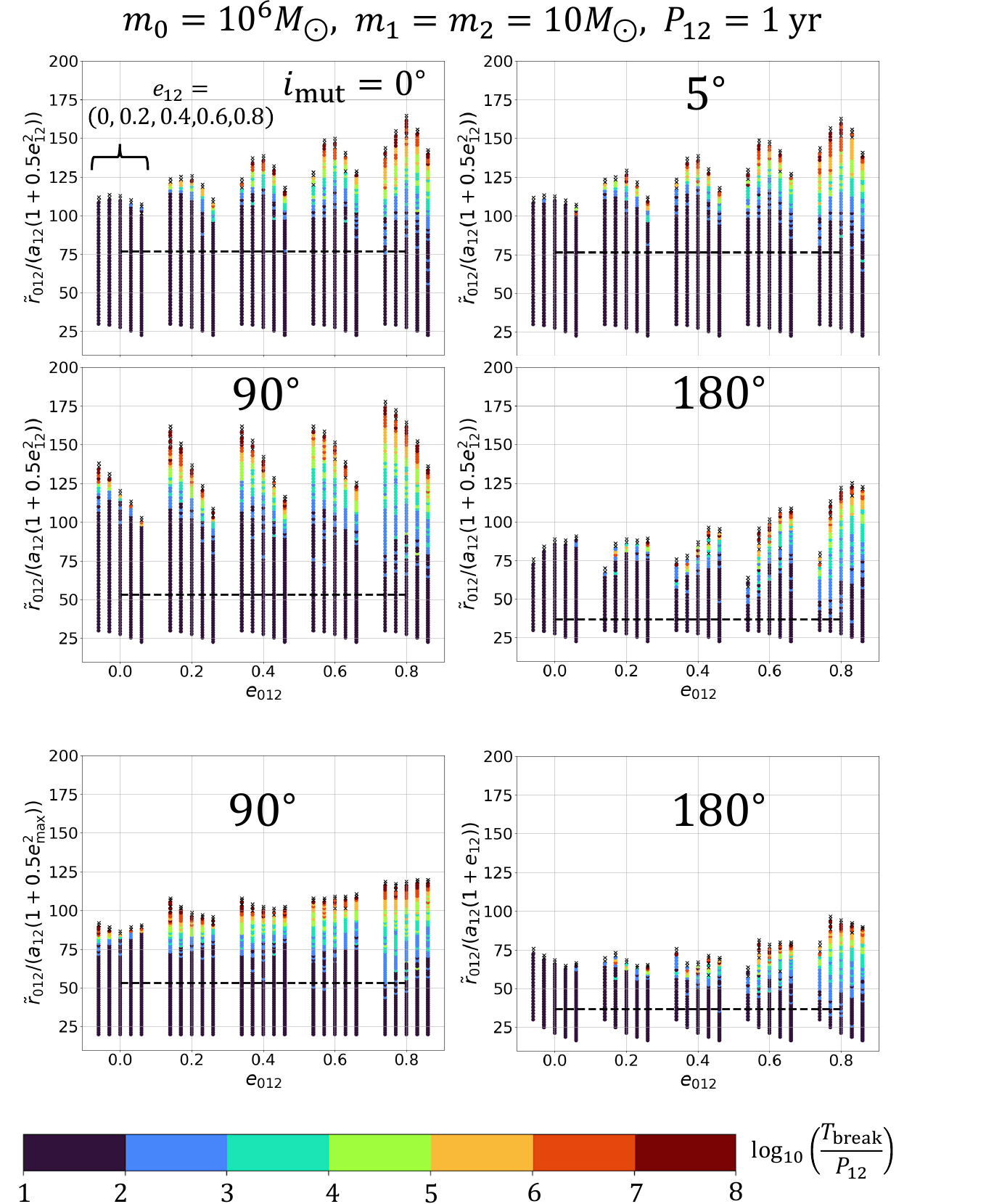}
\end{center}
\caption{Breakup time distribution in $e_\outs$-- distance ratio
  plane. Six panels show the results for $\imut=0^\circ$, $5^\circ$,
  $90^\circ$, and $180^\circ$. Each group of five represents the same
  value of $e_\outs$, but different values of $e_\ins$, shifted
  horizontally in the range of $0$, $0.2$, $0.4$, $0.6$ and $0.8$ from
  left to right. The upper four panels use the average distances of
  inner orbits, although lower
  two panels use different distance measures (see equation (\ref{eq:rin-tilde})). For references, we plot GPZM17 criterion (extrapolate $e_\outs=0$ results horizontally) by black dashed lines.
  \label{fig:imut}}
\end{figure*}

First, we show the results for four representative $\imut$
  values ($0^\circ$, $5^\circ$, $90^\circ$ and $180^\circ$), and
  explain how we derive the breakup condition formula. The
result is plotted in Figure \ref{fig:imut} in which the binary breakup
timescale of each system is indicated as a color-coded filled
circle. If a system does not break up within the integration time
limit $10^8 P_\ins$ (`unresolved' in \S \ref{sec:method}), the system is indicated by a cross symbol.  The
horizontal axis of the figure is $e_\outs$, and five sequences
centered at the value of $e_\outs$ correspond to the results that we
slightly shifted for visual clarity according to the value of $e_\ins$
($0$, $0.2$, $0.4$, $0.6$ and $0.8$ from left to right). For
reference, the dashed black lines in those panels indicate the GPZM17
criterion obtained for $e_\outs=0$.

The top four panels plot $a_\outs(1-e_\outs)/a_\ins(1+0.5e^2_\ins)$, the
pericenter distance between the binary and the central massive object
in units of the orbit-averaged distance of the binary, for different
$\imut$. While the effect of $e_\ins$ is reasonably absorbed in the
above scaling for $\imut=0^\circ$ and $5^\circ$, there remains a clear
systematic trend for $\imut=90^\circ$ and $180^\circ$.

Instead, we found that the residual $e_\ins$-dependence is well
absorbed by changing the vertical axis as
$a_\outs(1-e_\outs)/a_\ins(1+0.5e^2_\mathrm{max})$ for $\imut=90^\circ$
and $a_\outs(1-e_\outs)/a_\ins(1+e_\ins)$ for $\imut=180^\circ$; see the
bottom two panels in Figure \ref{fig:imut}.

The scaling with respect to $e_\ins$ for (near)-coplanar prograde
cases implies that the binary breakup due to the Hill-type instability
is not instantaneous in general, but occurs gradually. This is
reasonable especially for the large $T_\mathrm{break}/P_\ins$ results where
the outer orbital period is supposed to be much longer than the inner
orbital period as indicated from the conventional Hill stability
condition (\ref{eq:Hill-stability-Period}).

The scaling for $\imut=90^\circ$ was already suggested by GPZM17, and
can be understood as the orbit-averaged binary distance should be
computed from $e_\mathrm{max}$, instead of the initial value of
$e_\ins$, due to the ZKL oscillation.  We note that
\citet{Vynatheya2022} suggest that a similar approach works well for
HT-P systems.

The scaling for a coplanar retrograde system may be simply understood
  from the fact that the tidal interaction between the counter-orbiting
  two bodies is fairly instantaneous. Thus, it is determined by the
  configuration when the two bodies are the closest: when the outer orbit is at its pericenter $a_\outs(1-e_\outs)$, and
the inner orbit is at its apocenter $a_\ins(1+e_\ins)$.

As the result, we define $\tilde{r}_\outs\equiv a_\outs(1-e_\outs)$ and $\tilde{r}_\ins$ as a function of $\imut$, and model the breakup condition using $\tilde{r}_\outs/\tilde{r}_\ins$. In order to account for the $e_\ins$ dependence, we define $\tilde{r}_\ins$ based on the results in Figure \ref{fig:imut} as follows: $\tilde{r}_\ins = a_\ins (1+0.5e_\ins^2)$ for prograde systems without the ZKL oscillations, $\tilde{r}_\ins = a_\ins (1+0.5e_\mathrm{max}^2)$ for near-polar systems with the ZKL oscillations, and $\tilde{r}_\ins = a_\ins (1+e_\ins)$ for retrograde systems without the ZKL oscillations (see equation (\ref{eq:rin-tilde})).

\subsection{Outer eccentricity dependence
\label{subsec:eout}}
\begin{figure*}
	\begin{center}
		\includegraphics[clip,width=14cm]{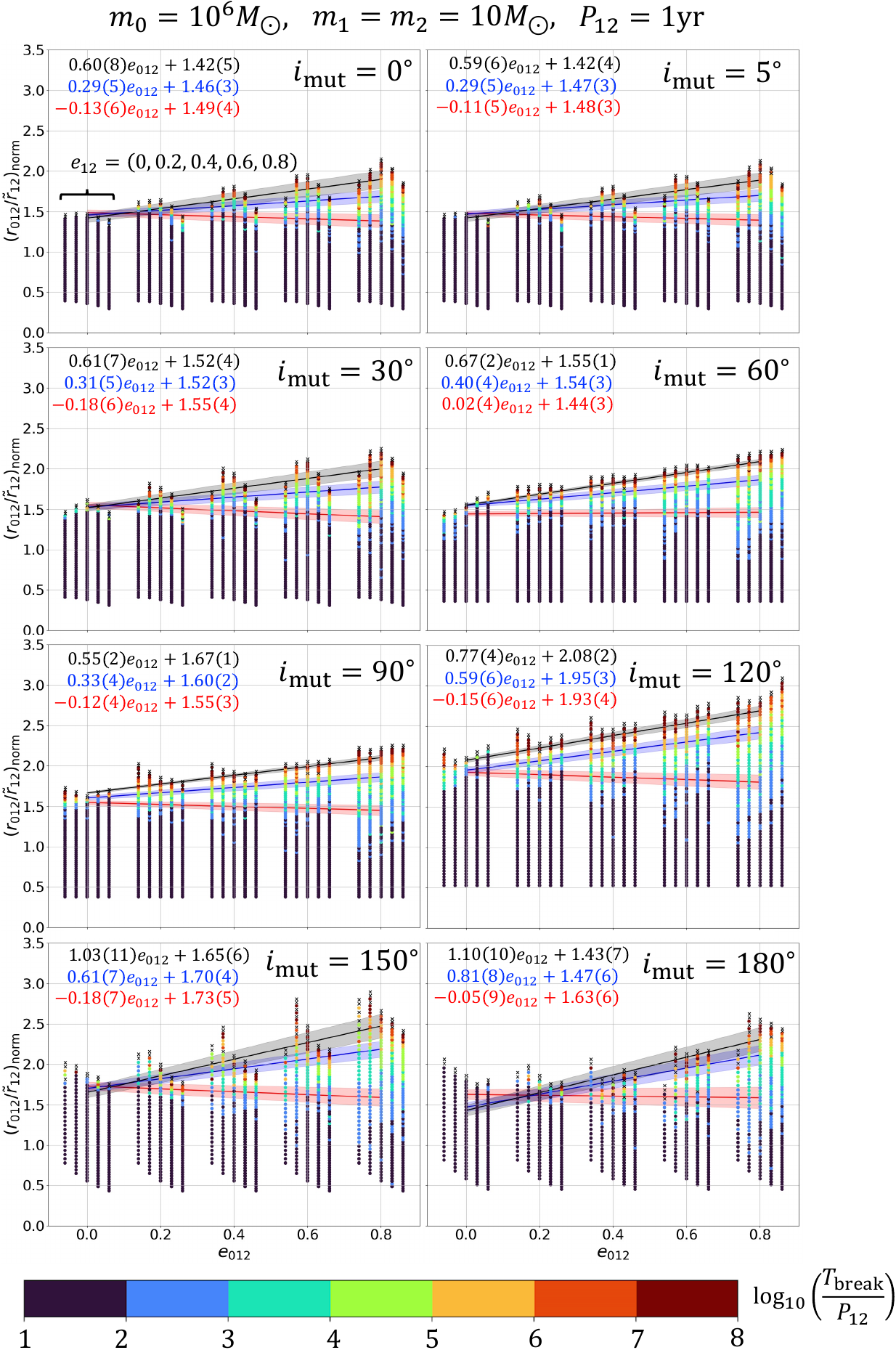}
	\end{center}
	\caption{Same as Figure \ref{fig:imut}, but the vertical
            axis is now using equation (\ref{eq:norm}). Red, blue, black lines are best-fit 
            results of $h(e_\outs, \imut; T_\mathrm{break}/P_\ins)$, corresponding to $T_\mathrm{break}/P_\ins=10^{2-4}$,
            $10^{4-6}$, and $10^{6-8}$, respectively. The shaded regions and upper-left texts of each panel are $1$--$\sigma$ errors and best-fit functions, respectively. \label{fig:imut_norm}}
\end{figure*}
Consider next the $e_\outs$-dependence of the stability boundary.  For
that purpose, we plot Figure \ref{fig:imut_norm} similarly to Figure \ref{fig:imut}, but in such a way that the vertical axis now adopts $\tilde{r}_\outs/\tilde{r}_\ins$ normalized by the GPZM17 criterion (\ref{eq:Grishin-criterion}) (see equation (\ref{eq:norm})). We
  classify systems into no ZKL--prograde, under ZKL, and no
  ZKL--retrograde systems, and use $\tilde{r}_\ins$ defined in the previous subsection (see  equation (\ref{eq:rin-tilde})). The resulting plot can be directly
  used to determine $h(e_\outs, \imut; T_\mathrm{break}/P_\ins)$ in
  our final expression (\ref{eq:Upsilon}).

The resulting Figure \ref{fig:imut_norm} shows the normalized $\tilde{r}_\outs/\tilde{r}_\ins$, and for comparison, we also plot the best-fit  $h(e_\outs, \imut; T_\mathrm{break}/P_\ins)$ with $1$--$\sigma$ errors by red ($T_\mathrm{break}=10^{2-4}P_\ins$), blue ($T_\mathrm{break}=10^{4-6}P_\ins$), and black ($T_\mathrm{break}=10^{6-8}P_\ins$) lines with shaded regions, respectively. We also include best-fit functions of $h(e_\outs, \imut; T_\mathrm{break}/P_\ins)$ in each panel with corresponding colored texts. Note that the values in parentheses are $1$--$\sigma$ errors, and assigned for final digits; for example, $0.60(8)$ and $1.03(11)$ mean $0.60 \pm 0.08$ and $1.03\pm 0.11$, respectively.

The best-fit $h(e_\outs=0, \imut; T_\mathrm{break}/P_\ins)$ functions are degenerate among different $T_\mathrm{break}/P_\ins$ for circular cases. This is partly because a binary quickly breaks even around the breakup boundary ($>10^8 P_\ins$) as stated later. Except for this behavior, the figure indicates that simple linear fits reasonably capture the $e_\outs$-dependence within the errors.

While our resulting $h(e_\outs, \imut; T_\mathrm{break}/P_\ins)$ is
systematically larger than that derived by GPZM17
(\ref{eq:Grishin-criterion}) corresponding to $h(e_\outs, \imut;
T_\mathrm{break}/P_\ins)=1$, it is partially explained by the
difference of the breakup definition (\S \ref{sec:method}) and our
longer integration time ($10^8~P_\ins$ instead of $100~P_\outs$ in
GPZM17). We also note that the results for 
  $T_\mathrm{break} = 10^{2-4}P_\ins$ (which is roughly equivalent to
  $10^2 P_\outs$ in our adopted models) is only weakly dependent on the value of $e_\outs$. Therefore, we conclude that our result is consistent with the GPZM17 result. We also note that the slope of $e_\outs$ in $h(e_\outs, \imut, T_\mathrm{break}/P_\ins =
    10^{6-8})$ is close to $0.6$ for near-coplanar prograde systems, which coincides with that of MA01 criterion for HT-P (see equation (\ref{eq:MA-criterion})) when
    $e_\outs \ll 1$.

The breakup timescale distribution in Figure \ref{fig:imut} (or Figure
\ref{fig:imut_norm}) also presents interesting features. For circular
outer orbits ($e_\outs=0$), especially for coplanar systems, a break
occurs suddenly with a very short timescale (typically $\lesssim
10P_\ins$), even around the breakup boundary ($>10^8 P_\ins$). In contrast,
  highly eccentric systems ($e_\outs > 0.4$) and polar systems ($\imut
  \sim 90^\circ$, in the strong ZKL oscillation regime) show gradually
  increasing breakup timescales towards the stability boundary at $10^8
  P_\ins$. This behavior is similar to the result in \citet{HTS2023}
  for HT-P systems, and indicates the importance to define the
  stability boundary as a function of the timescale for HT-S
  configurations as well. 
\subsection{Mass ratio dependence \label{subsec:massratio}}
\begin{figure*}
	\begin{center}
		\includegraphics[clip,width=14cm]{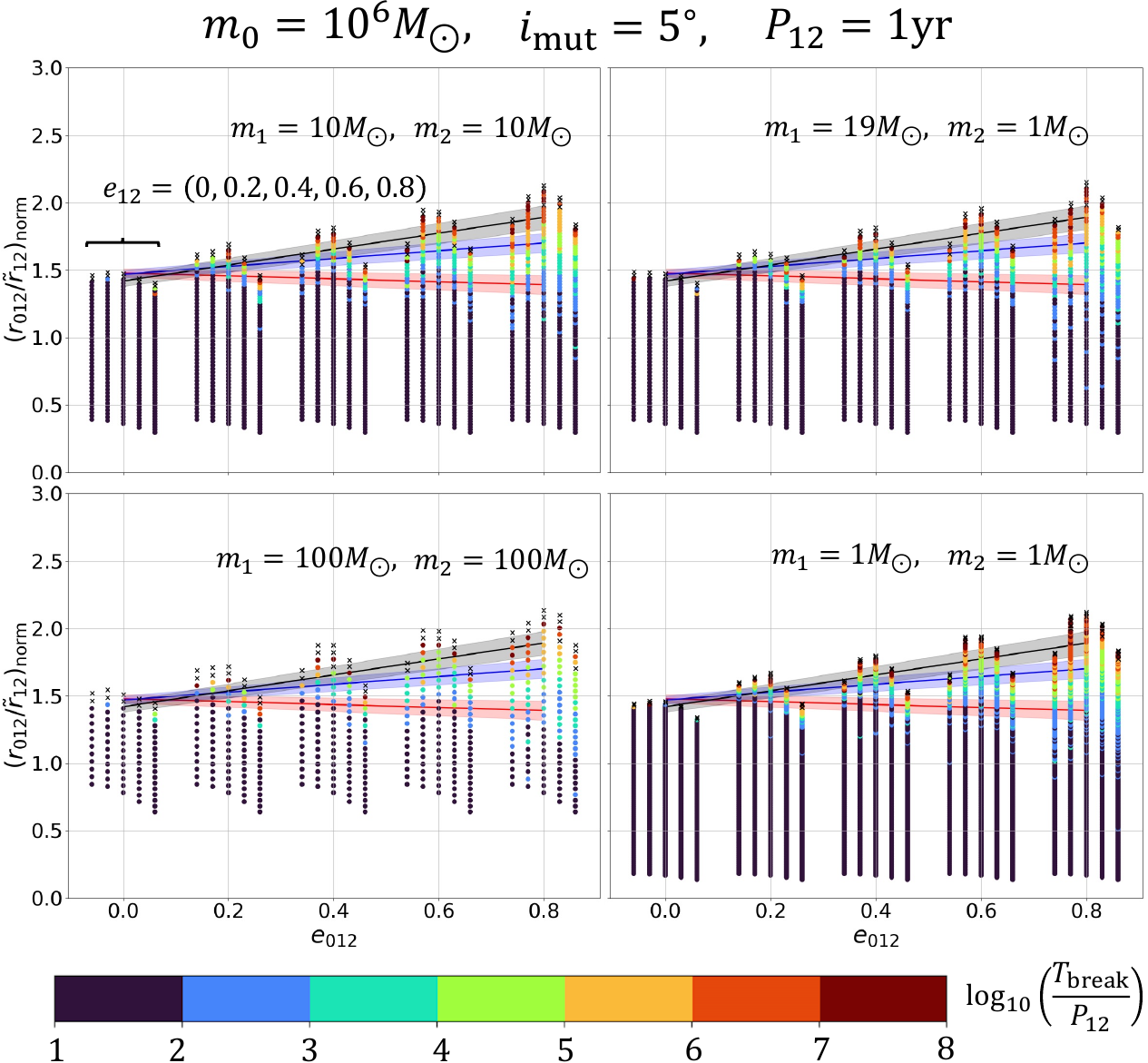}
	\end{center}
	\caption{Breakup time distribution on $\left(\tilde{r}_\outs/\tilde{r}_\ins\right)_\mathrm{norm}$ -- $e_\outs$
		plane. The vertical axis uses equation (\ref{eq:norm}) as in Figure \ref{fig:imut_norm}. Four panels correspond to different sets of $m_1$ and $m_2$, although fixing $m_0$ as $10^6~M_\odot$. The best-fit $h(e_\outs, \imut=5^\circ;T_\mathrm{break}/P_\ins)$ for $T_\mathrm{break}/P_\ins=10^{2-4}$ (red), $10^{4-6}$ (blue), $10^{6-8}$ (black) from Figure \ref{fig:imut_norm} are included as lines with error regions. \label{fig:mass}}
\end{figure*}
Figure \ref{fig:mass} shows how $T_\mathrm{break}/P_\ins$ distributions depend
  on the mass ratios, $m_{1}/m_0$ and $m_2/m_1$, in the cases of
  near-coplanar prograde systems ($\imut = 5^\circ$) with fixed
  initial phases.  While we fix  $m_0=10^6M_\odot$ and
  $P_\ins=1$ yr, the result are scalable with respect
  to those values; see equation (14) in \citet{HTS2022}. Upper-left,
  upper-right, lower-left and lower-right panels correspond to $(m_1,
  m_2)=(10M_\odot, 10M_\odot)$, $(19M_\odot, 1M_\odot)$, $(100M_\odot,
  100M_\odot)$, and $(1M_\odot, 1M_\odot)$, respectively. The
  vertical axis of Figure \ref{fig:mass} uses equation (\ref{eq:norm}) as we did in Figure \ref{fig:imut_norm}.

  Four panels in
  Figure \ref{fig:mass} appear to be almost identical, indicating that
  the mass dependence of the stability boundary is described by the
  conventional Hill-radius scaling $(m_0/ (m_1+m_2))^{1/3}$ alone,
  independent of $m_2/m_1$. For references, in Figure \ref{fig:mass}, we include best-fit $h(e_\outs, \imut=5^\circ; T_\mathrm{break}/P_\ins)$ from Figure \ref{fig:imut_norm}, as red ($T_\mathrm{break}=10^{2-4}P_\ins$), blue ($T_\mathrm{break}=10^{4-6}P_\ins$), and black ($T_\mathrm{break}=10^{6-8}P_\ins$) lines with error regions, respectively.

Finally, we also check the rescaling with mass, as expected from
Newtonian gravity, by varying $m_0$ values and fixing the mass ratios
$m_{1}/m_0$ and $m_2/m_1$. We confirm this rescaling property statistically, although chaotic behavior may change the result of individual simulations \citep[see][]{HTS2022}.
\section{Summary and conclusion \label{sec:summary}}

We have examined breakup condition and stability of
  hierarchical triple systems comprising a central massive body and a
  tight binary in eccentric orbits, which we referred to as HT-S
  systems. Using a series of direct N-body simulations, we empirically obtained the
  Hill-type stability criterion of HT-S systems, inequality (\ref{eq:Upsilon}),
  which generalizes the formula (\ref{eq:Grishin-criterion}) derived
  by \citet{Grishin2017}. 
  
\begin{figure*}
\begin{center}
\includegraphics[clip,width=13cm]{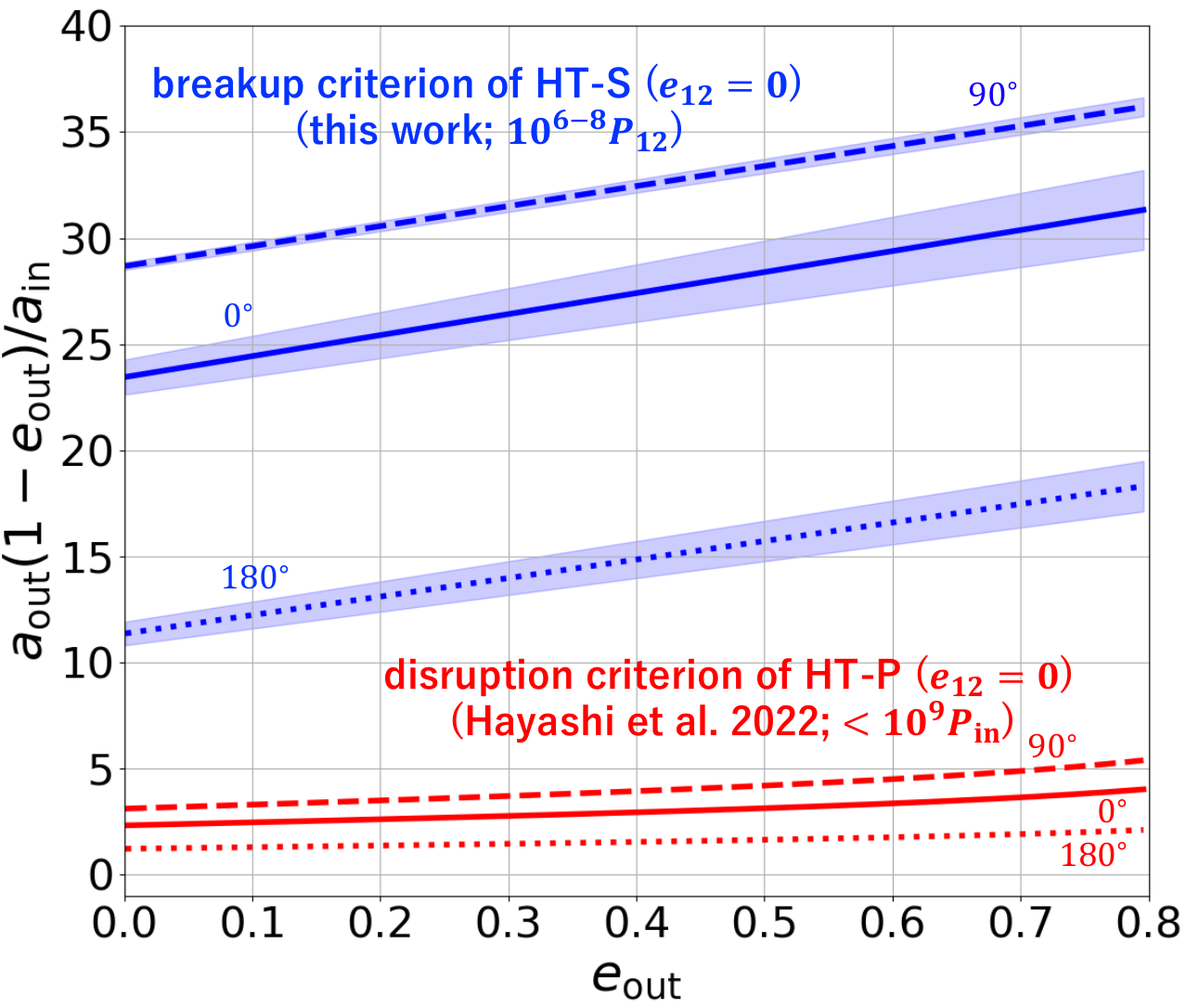}
\end{center}
\caption{Disruption criterion for HT-P (red curves, $e_\inp = 0$, $i_\mathrm{mut} = 0^\circ$, $90^\circ$, and $180^\circ$), and breakup criterion for HT-S
  (blue curves, $e_\ins=0$, $i_\mathrm{mut} = 0^\circ$, $90^\circ$, and $180^\circ$) in $e_\outp - a_\outp(1-e_\outp)/a_\inp$
  plane. We use equation (16) in \citet{HTS2022} for HT-P disruption criterion, and equation (\ref{eq:Upsilon}) adopting
    $T_\mathrm{break}/P_\ins = 10^{6-8}$ in this paper for HT-S
    breakup criterion, respectively. The errors of HT-S breakup criteria are plotted as blue regions.
  \label{fig:P-S-comparison2}}
\end{figure*}

We compare the binary breakup criterion (\ref{eq:Upsilon}) for
  HT-S systems and the disruption criterion for HT-P systems, equation
  (16) of \citet{HTS2022} in Figure \ref{fig:P-S-comparison2}. Here, we use `disruption' and `breakup' rather than just `stability' for definiteness. In the figure, HT-S breakup boundaries ($10^{6-8}P_\ins$) are plotted as blue lines (HT-S; this work), and HT-P disruption boundaries ($10^{9}P_\inp$) are included as red curves \citep[HT-P;][]{HTS2022}.  Note that the vertical axis of the figure is
  $a_\outp(1-e_\outp)/a_\mathrm{in}$ so as to plot both
  HT-S and HT-P systems simultaneously. Thus, the scaling with respect
  to $e_\ins$, equation (\ref{eq:rin-tilde}), is not
  incorporated in Figure \ref{fig:P-S-comparison2}.

Our major findings are summarized as follows.
\begin{description}
\item[(A) Outer eccentricity] The most important parameter that
  determines the stability of hierarchical triple systems is the ratio
  of the semi-major axes of the outer and inner orbits,
  $a_\outs/a_\ins$. In general, eccentricities of both orbits,
  $e_\outs$ and $e_\ins$, tend to destabilize the systems. The
  destabilizing effect of $e_\outs$ is mostly incorporated by
  replacing $a_\outs$ with $a_\outs(1-e_\outs)$, but the Hill-type stability boundaries 
  after the scaling still weakly increase as a function of
  $e_\outs$. We find that the residual effect for HT-S systems
    can be approximately captured by linear functions of $e_\outs$, at
    least for $e_\outs<0.8$ (see Figure \ref{fig:imut_norm} also), although its slope is sensitive to the choice of breakup timescales; within shorter breakup
    timescales, $e_\outs$ dependence becomes weaker.
\item[(B) Inner eccentricity] The effect of $e_\ins$ on the stability
   is weaker but more subtle than that of $e_\outs$. We find empirical
   scaling expressions for the $e_\ins$-dependence that is sensitive to
   $\imut$.
\item[(C) Mutual inclination of the inner and outer orbits] We
  adopted the $\imut$ dependence analytically derived in
  \citet{Grishin2017} following \citet{Innanen1980}. However, our
  simulations with $\imut = 30n^\circ$($n=0$--$6$), $5^\circ$
  find additional empirical corrections, as shown in Figures \ref{fig:fit} and \ref{fig:imut_norm}. 
\item[(D) Mass ratio] The mass dependence of the binary stability
   criterion is well described by the simple factor of
   $(m_0/(m_1+m_2))^{1/3}$, at least for near-coplanar prograde
   systems. It may be a bit surprising that the criterion is determined by
   the total mass of the initial binary $m_1+m_2$, and insensitive to
   their ratio $m_1/m_2$.
\end{description}

While we do not consider general relativity (GR) in the present work, the
Hill stability including GR effects becomes important for some
parameter ranges. For instance, \citet{Suzuki2020} derived a
criterion using analytic treatments concerning Sundman's inequality
and first-order post-Newtonian (1PN) approximate numerical simulations
for coplanar near-circular systems. They suggested that the PN
corrections tend to stabilize a system compared with a purely
Newtonian case. It is also well-known that the first-order GR
correction (apsidal precession term) suppresses the ZKL oscillations
if two timescales become comparable. Besides, gravitational wave
emissions reduce the energy for a long timescale, and therefore shrink
the semi-major axis of a binary. Close scatterings before collisions
may also be significantly affected by the GR corrections. Therefore,
the GR corrections affect the Hill stability of triple systems with
relevant parameter values.

The present paper focuses on the breakup condition of HT-S systems for definiteness. Nevertheless, the final fates of HT-S systems after binary breakups are also practically important to consider the whole dynamical evolution of HT-S systems, for instance a binary black hole orbiting around a supermassive black hole. As the first step, it will be important to study the relation between initial HT-S systems and resulting HT-P systems after breakups. This will be discussed elsewhere separately in the future. We also mention that the resulting HT-P systems may have much longer instability times than breakup timescales in general. Therefore, for the application to stellar triples, stellar evolution should also affect the stability and final fate of HT-S systems, as \citet {Toonen2022} point out for HT-P systems.

Finally, we would like to emphasize that the result in this paper is
basically scalable satisfying so called Kepler's third law, under the
assumption that a system is completely dominated by the Newtonian
gravity. Therefore, the result is applicable, not only for black hole
triples, but also for planetary (or satellite) systems.

\section*{Acknowledgments}

T.H. gratefully acknowledges Atsushi Taruya for fruitful discussions
about dynamics in three-body systems. The numerical simulations were
carried out on the local computer cluster \texttt{awamori}, and the
general calculation server from Center for Computational Astrophysics
(CfCA), National Astronomical Observatory of Japan (NAOJ).
T.H. gratefully acknowledges the fellowship by Japan Society for the
Promotion of Science (JSPS). This work is supported partly by the JSPS
KAKENHI grant No.  JP23K25908 (Y.S.), JP21J11378 and JP23KJ1153
(T.H.), and JP21K13914 (A.A.T.).

\appendix
\section{Effect of different initial phases on the breakup condition
  \label{appendix:initialphases}}
So far, all the simulation runs have been performed for a given
  set of initial phases of three bodies. We vary the initial phases so
  as to test if our breakup condition is robust against those
  changes.  For that purpose, we focus on coplanar prograde systems
  ($\imut=0^\circ$) in circular ($e_\ins=e_\outs=0$) and highly
  eccentric ($e_\ins=e_\outs=0.8$) orbits, and vary the initial phases
  as follows.

For circular systems, the pericenter arguments ($\omega_\ins,
\omega_\outs$) are irrelevant, and we arbitrarily fix them as $0^\circ$
that simply define the zero point of the locations of each body for
coplanar systems. In addition, due to symmetry, we set $M_\ins=0^\circ$
without loss of generality. Thus, the initial phases are parameterized
by the mean anomaly difference $\Delta M \equiv M_\outs - M_\ins=
M_\outs$ alone, and we vary it as $2^\circ i$ ($i= 1, 2, \cdots, 180$).

For eccentric systems, however, the initial phases are specified by
three independent parameters; the relative location of pericenters
($\Delta \omega \equiv \omega_\outs - \omega_\ins=\omega_\outs$), and mean anomalies
($M_\ins$ and $M_\outs$).  Therefore, we vary them as $\Delta \omega = \omega_\outs = 
30^\circ j$, $M_\ins=60^\circ k$, and $M_\outs=60^\circ l$, where the
integers $j$, $k$, and $l$ run from 1 to 6; see Figure
\ref{fig:orbit}.

Figure \ref{fig:phase1} plots how $T_\mathrm{break}/P_\ins$ for
circular systems is affected by the mean anomaly difference $\Delta
M$.  There is a clear systematic trend in the breakup boundary ($\sim 14$\% from $106$ to $122$), as a function of $\Delta M$. This result may seem
somewhat counter-intuitive, because the initial phases are expected to
be well mixed up after the orbital evolution.  However, the binary
breakup in the circular systems happens so quickly, $T_\mathrm{break} \lesssim
10P_\ins$, before the memory of the initial location is lost.

The variation of the stability boundary for circular systems turned out to be $\sim 14$\%. For comparison, we plot $h(e_\outs=0, \imut=0^\circ; T_\mathrm{break}/P_\ins=10^{2-4}, 10^{4-6}, 10^{6-8})$ range from Figure \ref{fig:imut_norm} as the orange region of Figure \ref{fig:phase1}. The result shows the errors of $h(e_\outs=0, \imut=0^\circ; T_\mathrm{break}/P_\ins=10^{2-4}, 10^{4-6}, 10^{6-8})$ reasonably cover the above systematic trend. While the origin of the above behavior
of circular systems may be possibly related to the chaos theory
\citep[e.g.][]{Mardling1995, Mardling1995b}, it is beyond the scope of
this paper and we do not consider further.

\begin{figure*}[h]
	\begin{center}
		\includegraphics[clip,width=16cm]{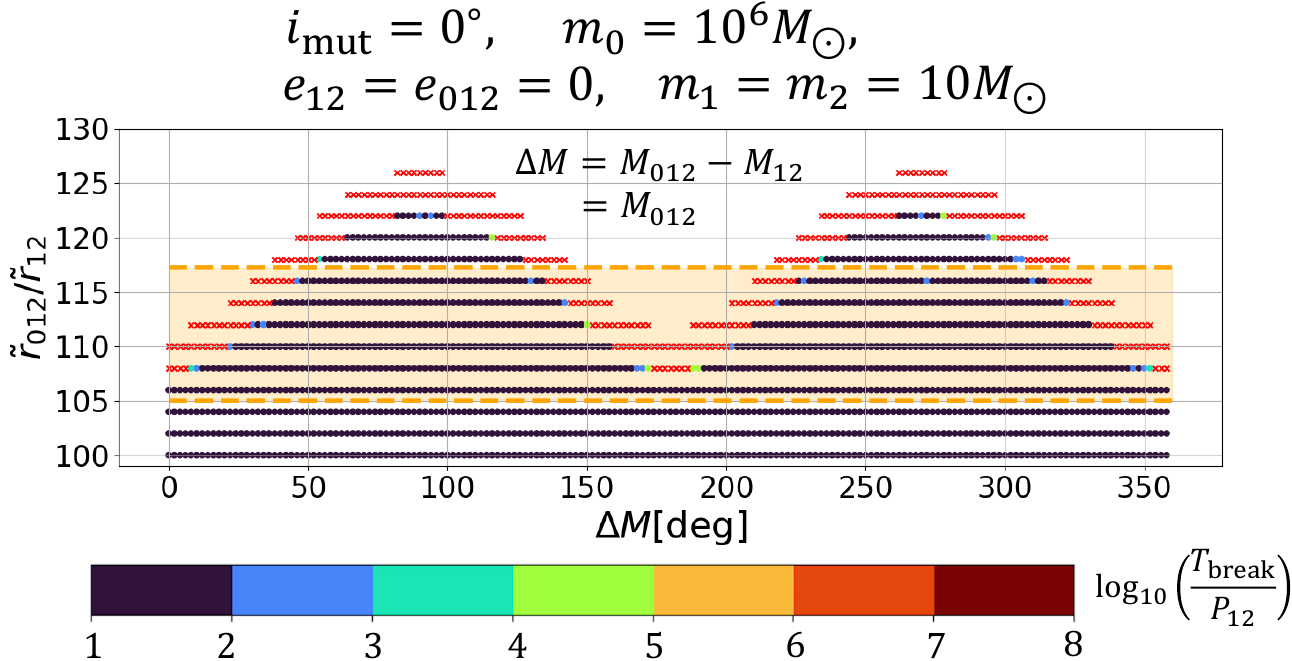}
	\end{center}
	\caption{Breakup time distribution in $\Delta M$-- $\tilde{r}_\outs / \tilde{r}_\ins$ plane. The red crosses denote `unresolved' systems. We plot $h(e_\outs=0, \imut=0^\circ; T_\mathrm{break}/P_\ins = 10^{2-4},10^{4-6}, 10^{6-8})$ ranges from Figure \ref{fig:imut_norm} with orange region.\label{fig:phase1}}
\end{figure*}
\begin{figure*}
	\begin{center}
		\includegraphics[clip,width=14cm]{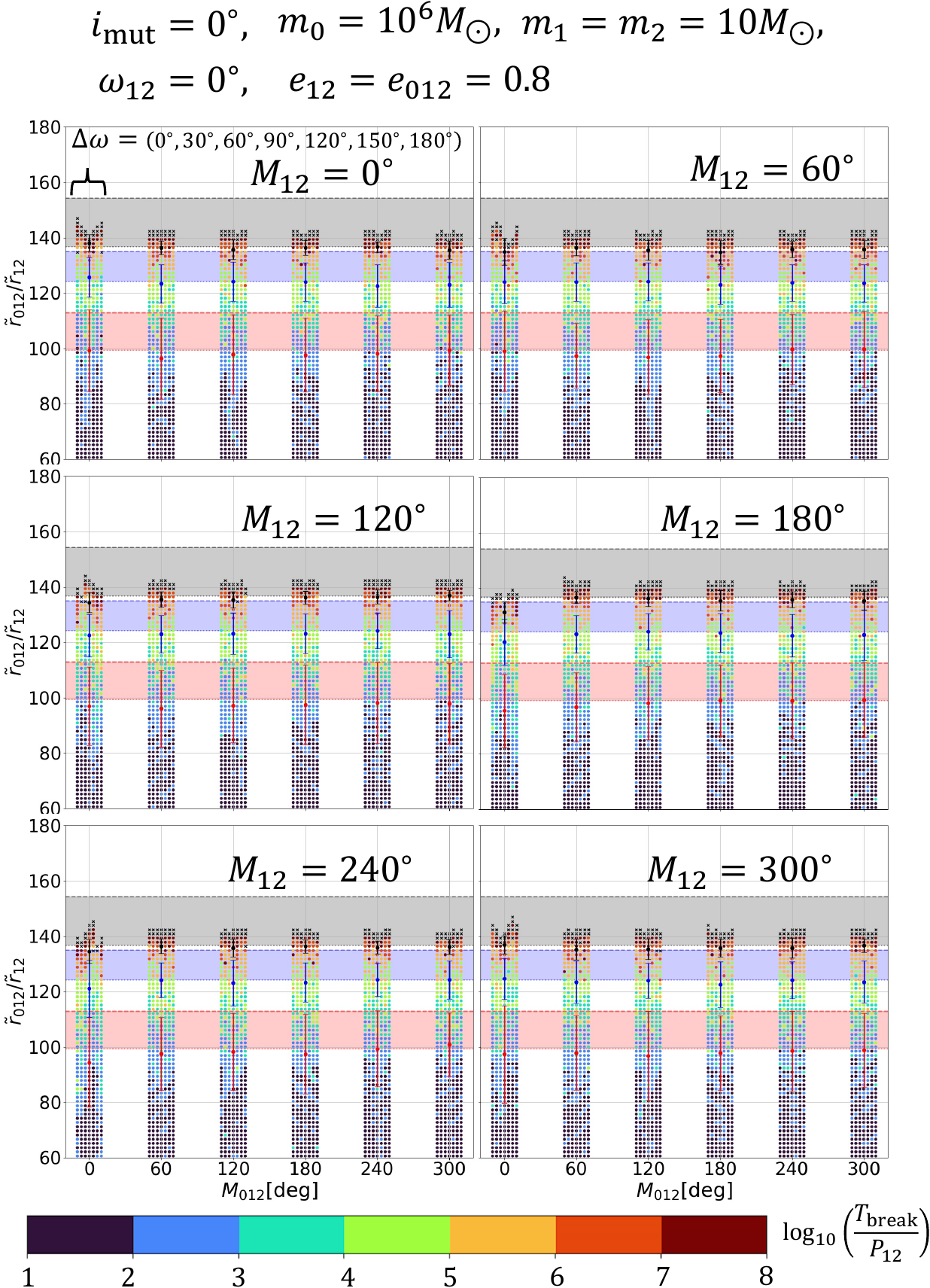}
	\end{center}
	\caption{Breakup time distribution in $M_\outs$ --
		$\tilde{r}_\outs/\tilde{r}_\ins$ plane. We include best-fit $h(e_\outs, \imut; T_\mathrm{break}/P_\ins)$ from Figure \ref{fig:imut_norm} with errors by shaded regions. We plot means and standard deviations of $\tilde{r}_\outs/\tilde{r}_\ins$ as colored circles with error bars. Red, blue, and black colors correspond to $T_\mathrm{break}/P_\ins = 10^{2-4}$, $10^{4-6}$, $10^{6-8}$, respectively. 
		\label{fig:phase2}}
\end{figure*}
Figure \ref{fig:phase2} shows the results for eccentric systems. Each
panel corresponds to different values of $M_\ins$, and the horizontal
axis denotes $M_\outs$ here. Similarly to the visualization treatment
for $e_\ins$ in the previous sections, we plot the results for
different $\Delta \omega$ by shifting data sequences
horizontally. For references, in Figure  \ref{fig:phase2}, we include the best-fit $h(e_\outs=0.8, \imut=0^\circ, T_\mathrm{break}/P_\ins)$ ranges from Figure \ref{fig:imut_norm} as red ($T_\mathrm{break}=10^{2-4}P_\ins$), blue ($T_\mathrm{break}=10^{4-6}P_\ins$), and black regions ($T_\mathrm{break}=10^{6-8}P_\ins$), respectively.

In Figure \ref{fig:phase2}, we also plot the means and standard deviations of $\tilde{r}_\outs/\tilde{r}_\ins$ as colored circles with error bars for comparison; red, blue, and black colors correspond to $T_\mathrm{break}/P_\ins=10^{2-4}$, $10^{4-6}$, and $10^{6-8}$, respectively. They are calculated for the binned data, following the same binning method as fitting in \S \ref{sec:result} (see the paragraph above equation (\ref{eq:norm})). The result clearly shows that the means of $\tilde{r}_\outs/\tilde{r}_\ins$ have no systematic initial phase dependence, and they are basically consistent with $h(e_\outs=0.8, \imut=0^\circ, T_\mathrm{break}/P_\ins)$.

Contrary to the circular systems, Figure \ref{fig:phase2}
shows no systematic dependence on initial phases for eccentric
systems. This is because eccentric systems around the breakup boundary are relatively more stable and
the longer timescale for the orbital evolution before the instability
loses the memory of their initial phases. So, we conclude that the initial phase dependence is negligible for breakup boundaries of the eccentric systems.



\begin{thebibliography}{}
	\expandafter\ifx\csname natexlab\endcsname\relax\def\natexlab#1{#1}\fi
	\providecommand{\url}[1]{\href{#1}{#1}}
	\providecommand{\dodoi}[1]{doi:~\href{http://doi.org/#1}{\nolinkurl{#1}}}
	\providecommand{\doeprint}[1]{\href{http://ascl.net/#1}{\nolinkurl{http://ascl.net/#1}}}
	\providecommand{\doarXiv}[1]{\href{https://arxiv.org/abs/#1}{\nolinkurl{https://arxiv.org/abs/#1}}}
	
	\bibitem[{{Eggleton} \& {Kiseleva}(1995)}]{Eggleton1995}
	{Eggleton}, P., \& {Kiseleva}, L. 1995, \apj, 455, 640, \dodoi{10.1086/176611}
	
	\bibitem[{{Georgakarakos}(2013)}]{Georgakarakos2013}
	{Georgakarakos}, N. 2013, \na, 23, 41, \dodoi{10.1016/j.newast.2013.02.004}
	
	\bibitem[{{Grishin} {et~al.}(2017){Grishin}, {Perets}, {Zenati}, \&
		{Michaely}}]{Grishin2017}
	{Grishin}, E., {Perets}, H.~B., {Zenati}, Y., \& {Michaely}, E. 2017, \mnras,
	466, 276, \dodoi{10.1093/mnras/stw3096}
	
	\bibitem[{{Hayashi} {et~al.}(2022){Hayashi}, {Trani}, \& {Suto}}]{HTS2022}
	{Hayashi}, T., {Trani}, A.~A., \& {Suto}, Y. 2022, \apj, 939, 81,
	\dodoi{10.3847/1538-4357/ac8f48}
	
	\bibitem[{{Hayashi} {et~al.}(2023){Hayashi}, {Trani}, \& {Suto}}]{HTS2023}
	---. 2023, \apj, 943, 58, \dodoi{10.3847/1538-4357/acac1e}
	
	\bibitem[{{He} \& {Petrovich}(2018)}]{He2018}
	{He}, M.~Y., \& {Petrovich}, C. 2018, \mnras, 474, 20,
	\dodoi{10.1093/mnras/stx2718}
	
	\bibitem[{Hill(1878)}]{Hill1878}
	Hill, G.~W. 1878, American Journal of Mathematics, 1, 5.
	\newblock \url{http://www.jstor.org/stable/2369430}
	
	\bibitem[{{Holman} \& {Wiegert}(1999)}]{Holman1999}
	{Holman}, M.~J., \& {Wiegert}, P.~A. 1999, \aj, 117, 621,
	\dodoi{10.1086/300695}
	
	\bibitem[{{Innanen}(1980)}]{Innanen1980}
	{Innanen}, K.~A. 1980, \aj, 85, 81, \dodoi{10.1086/112642}
	
	\bibitem[{{Kozai}(1962)}]{Kozai1962}
	{Kozai}, Y. 1962, \aj, 67, 591, \dodoi{10.1086/108790}
	
	\bibitem[{{Lalande} \& {Trani}(2022)}]{Lalande2022}
	{Lalande}, F., \& {Trani}, A.~A. 2022, \apj, 938, 18,
	\dodoi{10.3847/1538-4357/ac8eab}
	
	\bibitem[{{Lewis} {et~al.}(2015){Lewis}, {Ochiai}, {Nagasawa}, \&
		{Ida}}]{Lewis2015}
	{Lewis}, K.~M., {Ochiai}, H., {Nagasawa}, M., \& {Ida}, S. 2015, \apj, 805, 27,
	\dodoi{10.1088/0004-637X/805/1/27}
	
	\bibitem[{{Li} {et~al.}(2015){Li}, {Naoz}, {Kocsis}, \& {Loeb}}]{Li2015}
	{Li}, G., {Naoz}, S., {Kocsis}, B., \& {Loeb}, A. 2015, \mnras, 451, 1341,
	\dodoi{10.1093/mnras/stv1031}
	
	\bibitem[{{Lidov}(1962)}]{Lidov1962}
	{Lidov}, M.~L. 1962, \planss, 9, 719, \dodoi{10.1016/0032-0633(62)90129-0}
	
	\bibitem[{{Marchal} \& {Bozis}(1982)}]{Marchal1982}
	{Marchal}, C., \& {Bozis}, G. 1982, Celestial Mechanics, 26, 311,
	\dodoi{10.1007/BF01230725}
	
	\bibitem[{{Mardling} \& {Aarseth}(1999)}]{Mardling1999}
	{Mardling}, R., \& {Aarseth}, S. 1999, in NATO Advanced Science Institutes
	(ASI) Series C, Vol. 522, NATO Advanced Science Institutes (ASI) Series C,
	ed. B.~A. {Steves} \& A.~E. {Roy} (Springer), 385
	
	\bibitem[{{Mardling}(1995{\natexlab{a}})}]{Mardling1995}
	{Mardling}, R.~A. 1995{\natexlab{a}}, \apj, 450, 722, \dodoi{10.1086/176178}
	
	\bibitem[{{Mardling}(1995{\natexlab{b}})}]{Mardling1995b}
	---. 1995{\natexlab{b}}, \apj, 450, 732, \dodoi{10.1086/176179}
	
	\bibitem[{{Mardling} \& {Aarseth}(2001)}]{Mardling2001}
	{Mardling}, R.~A., \& {Aarseth}, S.~J. 2001, \mnras, 321, 398,
	\dodoi{10.1046/j.1365-8711.2001.03974.x}
	
	\bibitem[{{Myll{\"a}ri} {et~al.}(2018){Myll{\"a}ri}, {Valtonen}, {Pasechnik},
		\& {Mikkola}}]{Myllari2018}
	{Myll{\"a}ri}, A., {Valtonen}, M., {Pasechnik}, A., \& {Mikkola}, S. 2018,
	\mnras, 476, 830, \dodoi{10.1093/mnras/sty237}
	
	\bibitem[{{Ochiai} {et~al.}(2014){Ochiai}, {Nagasawa}, \& {Ida}}]{Ochiai2014}
	{Ochiai}, H., {Nagasawa}, M., \& {Ida}, S. 2014, \apj, 790, 92,
	\dodoi{10.1088/0004-637X/790/2/92}
	
	\bibitem[{{Suzuki} {et~al.}(2020){Suzuki}, {Nakamura}, \&
		{Yamada}}]{Suzuki2020}
	{Suzuki}, H., {Nakamura}, Y., \& {Yamada}, S. 2020, \prd, 102, 124063,
	\dodoi{10.1103/PhysRevD.102.124063}
	
	\bibitem[{{Toonen} {et~al.}(2022){Toonen}, {Boekholt}, \& {Portegies
			Zwart}}]{Toonen2022}
	{Toonen}, S., {Boekholt}, T.~C.~N., \& {Portegies Zwart}, S. 2022, \aap, 661,
	A61, \dodoi{10.1051/0004-6361/202141991}
	
	\bibitem[{{Tory} {et~al.}(2022){Tory}, {Grishin}, \& {Mandel}}]{Tory2022}
	{Tory}, M., {Grishin}, E., \& {Mandel}, I. 2022, \pasa, 39, e062,
	\dodoi{10.1017/pasa.2022.57}
	
	\bibitem[{{Trani} \& {Spera}(2023)}]{Trani2023}
	{Trani}, A.~A., \& {Spera}, M. 2023, IAU Symposium, 362, 404,
	\dodoi{10.1017/S1743921322001818}
	
	\bibitem[{{VanLandingham} {et~al.}(2016){VanLandingham}, {Miller}, {Hamilton},
		\& {Richardson}}]{VanLandingham2016}
	{VanLandingham}, J.~H., {Miller}, M.~C., {Hamilton}, D.~P., \& {Richardson},
	D.~C. 2016, \apj, 828, 77, \dodoi{10.3847/0004-637X/828/2/77}
	
	\bibitem[{{von Zeipel}(1910)}]{Zeipel1910}
	{von Zeipel}, H. 1910, Astronomische Nachrichten, 183, 345,
	\dodoi{10.1002/asna.19091832202}
	
	\bibitem[{{Vynatheya} {et~al.}(2022){Vynatheya}, {Hamers}, {Mardling}, \&
		{Bellinger}}]{Vynatheya2022}
	{Vynatheya}, P., {Hamers}, A.~S., {Mardling}, R.~A., \& {Bellinger}, E.~P.
	2022, \mnras, 516, 4146, \dodoi{10.1093/mnras/stac2540}
	
	\bibitem[{{Wei} {et~al.}(2021){Wei}, {Naoz}, {Faridani}, \& {Farr}}]{Wei2021}
	{Wei}, L., {Naoz}, S., {Faridani}, T., \& {Farr}, W.~M. 2021, \apj, 923, 118,
	\dodoi{10.3847/1538-4357/ac2c70}
	
	\bibitem[{{Xuan} {et~al.}(2023){Xuan}, {Naoz}, \& {Chen}}]{Xuan2023}
	{Xuan}, Z., {Naoz}, S., \& {Chen}, X. 2023, \prd, 107, 043009,
	\dodoi{10.1103/PhysRevD.107.043009}
	
	\bibitem[{{Zhang} {et~al.}(2023){Zhang}, {Naoz}, \& {Will}}]{Zhang2023}
	{Zhang}, E., {Naoz}, S., \& {Will}, C.~M. 2023, \apj, 952, 103,
	\dodoi{10.3847/1538-4357/acd782}
	
\end{thebibliography}
\end{document}